\PassOptionsToPackage{table,svgnames,xcdraw}{xcolor}

\newif\ifccs

\ifccs

\documentclass[sigconf]{acmart}
\AtBeginDocument{%
  }

\copyrightyear{2024}
\acmYear{2024}
\setcopyright{rightsretained}
\acmConference[WWW '24 Companion]{Companion Proceedings of the ACM Web Conference 2024}{May 13--17, 2024}{Singapore, Singapore}
\acmBooktitle{Companion Proceedings of the ACM Web Conference 2024 (WWW '24 Companion), May 13--17, 2024, Singapore, Singapore}\acmDOI{10.1145/3589335.3651960}
\acmISBN{979-8-4007-0172-6/24/05}

\makeatletter
\gdef\@copyrightpermission{
  \begin{minipage}{0.3\columnwidth}
   \href{https://creativecommons.org/licenses/by/4.0/}{\includegraphics[width=0.90\textwidth]{4ACM-CC-by-88x31.eps}}
  \end{minipage}\hfill
  \begin{minipage}{0.7\columnwidth}
   \href{https://creativecommons.org/licenses/by/4.0/}{This work is licensed under a Creative Commons Attribution International 4.0 License.}
  \end{minipage}
  \vspace{5pt}
}
\makeatother

\else

\documentclass[10pt,a4paper]{article}
\usepackage[utf8]{inputenc}

\makeatother
\usepackage[hidelinks, bookmarksopen=true]{hyperref}
\usepackage{bookmark}

\fi

\hypersetup{
    colorlinks,
    linkcolor={red!50!black},
    citecolor={red!50!black},
    urlcolor={blue!80!black}
}
\usepackage{graphicx} 
\usepackage{amsfonts} 
\usepackage{tikz}
\usepackage{booktabs}
\usepackage[super]{nth}
\usepackage{adjustbox,lipsum}
\usepackage{pgfplots}
\usepackage{lipsum}
\usepackage{dblfloatfix}
\usepackage{array}
\usepackage{amsmath}
\usepackage{listings}
\usepackage{colortbl}
\usepackage{makecell}

\ifccs
\fi

\definecolor{listgray}{rgb}{0.88,0.88,0.88} 
\usepackage{booktabs,makecell,multirow,tabularx}

\newcommand{\eg}{e.g., }
\newcommand{\ie}{i.e., }

\begin{document}
\definecolor{marycolor}{rgb}{1,.8,.79}

\title{Statistical Confidence in Mining Power Estimates for PoW Blockchains}

\ifccs
\else

\author{
Mary Milad \\ University of Edinburgh \\ mmilad@ed.ac.uk
\and Christina Ovezik \\ University of Edinburgh \\ christina.ovezik@ed.ac.uk
\and Dimitris Karakostas \\ University of Edinburgh \\ d.karakostas@ed.ac.uk
\and Daniel W. Woods \\ University of Edinburgh \\ daniel.woods@ed.ac.uk
}

\fi

\ifccs
\else

\maketitle

\fi

\begin{abstract}
The security of blockchain systems depends on the distribution of mining power across participants.
If sufficient mining power is controlled by one entity, they can force their own version of events. This may allow them to double spend coins, for example. 
For Proof of Work (PoW) blockchains, however, the distribution of mining power cannot be read directly from the blockchain and must instead be inferred from the number of blocks mined in a specific sample window. 
We introduce a framework to quantify this statistical uncertainty for the Nakamoto coefficient, which is a commonly-used measure of blockchain decentralization.
We show that aggregating blocks over a day can lead to considerable uncertainty, with Bitcoin failing more than half the hypothesis tests ($\alpha=0.05$) when using a daily granularity.
For these reasons, we recommend that blocks are aggregated over a sample window of at least 7 days.
Instead of reporting a single value, our approach produces a range of possible Nakamoto coefficient values that have statistical support at a particular significance level $\alpha$.

\end{abstract}

\section{Introduction}
The security of Proof of Work (PoW) blockchains depends on the distribution of mining power across participants~\cite{EC:GarKiaLeo15, zhang2019security, narayanan2016bitcoin}.
For example, an attacker controlling the majority of the mining power in the Bitcoin network could rewrite the blockchain and double spend coins (aptly termed a 51\% attack).
However, unlike, say, stake in a Proof of Stake (PoS) system, the exact mining power of each party cannot be directly retrieved from PoW blockchains.

State-of-the-art approaches to estimating mining power rely on the fact that an entity with X\% of the mining power has an X\% chance of mining a specific block~\cite{sok_stratified}.
Therefore, entities with more mining power are more likely to have mined more blocks.
By the same reasoning, entities who mined more blocks in a time period likely held more mining power. 
Indeed, various authors \cite{measuringdec, evolution, individual, beikverdi2015trend} assume that
an entity who mined X\% of the blocks in a certain period held X\% of the mining power.

This inference is imperfect though.
A trivial example illustrates what this assumption depends upon, namely observing sufficient blocks. 
If calculations are based on observing just one block, the above approach would infer that the entity who mined that block holds $100\%$ of the mining power.
In reality, this miner was just the winner of a single stochastic process, and likely holds a fraction of the mining share.
Clearly the solution is to observe more than one block, but the question is, how many blocks do we need to observe to obtain reliable estimates? This is a classic problem in statistical inference.

We answer this question with respect to a metric that summarizes the distribution of mining power, namely the Nakamoto coefficient (NC). The Nakamoto coefficient is defined as the minimum number of independent entities needed to obtain the majority ownership of a blockchain \cite{quantifying}. In these settings, the Nakamoto coefficient measures the resilience of a network by capturing blockchain decentralization, yet there is no method for calculating whether \emph{enough} blocks have been observed.

This paper provides an approach to quantifying statistical uncertainty when determining the Nakamoto coefficient. We answer the following research questions:

\textbf{RQ1} How can we evaluate statistical confidence in NC estimates?\\ 
\indent\textbf{RQ2} Are estimates in prior work statistically sound?
\\
\textbf{Contribution}
We introduce a framework that models the probability of a group of entities mining a block as a binomial distribution.
This approach enables us to calculate whether a given Nakamoto coefficient estimate passes a hypothesis test at a given confidence level $\alpha$.
We show that most estimates for Bitcoin do not pass a binomial hypothesis test with $\alpha=0.05$ if a daily granularity is used.
The framework also provides a range of plausible values for the Nakamoto coefficient, which have statistical support.
We also provide Python code extracts (Appendix~\ref{appendix: code}) using our framework to support future work.\footnote{The full code that we used, as well as some sample data, can be found on this public GitHub repository: \url{https://github.com/Blockchain-Technology-Lab/nc-statistical-confidence}.}

Section~\ref{sec:approach} introduces our statistical framework and data collection choices.
Section~\ref{sec:empirical results} displays the statistical confidence in Nakamoto coefficient estimates across various blockchains.
Section~\ref{sec:litreview} identifies the methodological choices in prior work.
Section~\ref{sec:discussion} discusses our framework and results, while Section~\ref{sec:conclusion} offers a conclusion.


\section{Approach} \label{sec:approach}
We introduce a framework to evaluate the statistical uncertainty in estimates of mining power distribution in Section~\ref{subsec: conceptual framework}.
We then explain how we collected empirical data in Section ~\ref{subsec: data collection}.

\subsection{Framework}
\label{subsec: conceptual framework}
To address \textbf{RQ1}, we develop methods to evaluate and visualize statistical confidence in the Nakamoto coefficient at the consensus layer for PoW blockchains. At a point in time, the Nakamoto coefficient can be modelled as a multinomial distribution in which each miner holds a proportion, $p_i$, of the total mining power where $\sum_{i=1}^np_i = 1$. 
A Nakamoto coefficient of $C$ is equivalent to saying $C$ is the smallest integer such that $\sum_{i=1}^C p_i > 0.5$ where $p_1$, ... $p_n$ are ordered in decreasing size. 

The proportion of mining power ($p_i$) owned by entity $i$, cannot be directly observed, and must instead be inferred from the share of blocks mined by each entity.
The $i$-th entity is expected to mine $n \times p_i$ blocks given $n$ opportunities.
Due to the stochastic mining process, the actual number of blocks mined are randomly distributed around this expected value.
By the law of large numbers, the proportion of blocks $\hat{p_i}$ mined per entity converges on $p_i$ given enough observations $n$.
Thus, we can use the observed proportions $\hat{p_i}$ as a proxy for the true mining power $p_i$ providing we accept statistical uncertainty.

Turning to the Nakamoto coefficient, the question is not whether the $i$-th miner holds $p_i$ of the mining power, but instead whether the top $C$ miners collectively hold mining power $p_C > 0.5$.
This reduces the mulitnomial distribution to a simple binomial distribution in which $n$ is the number of blocks in a sample and $k$ is the number of blocks mined by the top $C$ miners, such that $\hat{p_C}=\frac{k}{n}$.
This allows us to conduct a hypothesis test of a Nakamoto coefficient estimate at a given significance level $\alpha$, as follows:

\begin{center}
$H_0: p_C \leq 0.5$
\\
$H_1: p_C > 0.5$
\end{center}
We assume the underlying distribution is a binomial $B(0.5)$, and reject $H_0$ if the probability of observing $\hat{p_C}$ is less than $\alpha$.
If the blockchain is such that an attack can be launched if an entity controls $X\%$ of the mining power, then the $p$-values should be calculated according to $B(X)$.
If the null hypothesis is rejected, then there is statistical support for saying that the true Nakamoto coefficient is  $C$ or less. 

The significance level $\alpha$ can be viewed as the acceptable rate of false positives.
It is the rate at which the null hypothesis is rejected when it is in fact true.
An $\alpha$ value of $0.05$ is the most common choice within the scientific community, while lower values can be used to provide higher confidence in the results~\cite{biau2010p}.

Note that, if the null hypothesis is rejected for some $C$, then the Nakamoto coefficient is not necessarily equal to $C$, as the top $C-1$ miners may also have mined enough blocks in the $n$ trials to reject $H_0$.
This motivates the concept of a confidence interval $[a, b]$ of possible Nakamoto coefficients at the $\alpha$ confidence level. 
This range is such that $a$ is the smallest number of top miners for which the null hypothesis would be rejected, and b is the largest value of top miners for which the alternate hypothesis would be rejected. These can be seen as the upper and lower bounds for the Nakamoto coefficient.

If a given observation of $n$ blocks has a confidence window $[a, b]$ such that $a < b$, then we should report the Nakamoto coefficient as a range, since we cannot reject any of those values at the significance level $\alpha$.
This allows us to visualize the statistical uncertainty in a given estimate, and the $p$-values allow us to quantify uncertainty.
The range is analogous to an error bar.

Through hypothesis tests using real blockchain data, we will show that the existing approach to calculating the Nakamoto coefficient create a false sense of security by obscuring the statistical uncertainty.
These empirical results can be found in Section \ref{sec:empirical results}.

\subsection{Data Collection}
\label{subsec: data collection}

\begin{table} 
\caption{Data we collected using Google BigQuery.}
\label{table:data information}
\centering
\adjustbox{max width=\textwidth}{
\begin{tabular}{ccccccc} 
\toprule
Ledger & Sample Window & Mean blocks/day \\
\midrule
Bitcoin & Jan '18 - Aug '23 & 146 \\ 
Bitcoin Cash & Jan '18 - Nov '23 & 143 \\ 
Ethereum & Jan '18 - September '22 &   6229 \\
Litecoin & Jan' 18 - Nov '23 &  577\\
Zcash & Jan '18 - May '22 &  891 \\
\bottomrule
\end{tabular}}
\end{table}

To apply this framework, we first collected data about the distribution of produced blocks of 5 different PoW blockchains, using datasets provided by BigQuery~\cite{fernandes2015bigquery}. Table~\ref{table:data information} shows the ledgers that we analyzed, together with the sample window we used and the mean number of blocks per day for each of them.\footnote{Data on Ethereum is restricted to prior to the transition to Proof of Stake which occurred on September 15, 2022.} The raw data from BigQuery was manually parsed such that each entry corresponds to a block and contains the block's number, its timestamp, the address which receives the fees from the block
, and some identifiers that can potentially be used to attribute the block to the entity that produced it.\footnote{For Bitcoin, Bitcoin Cash, Litecoin and Zcash, this corresponds to the \emph{coinbase\_param} field from the BigQuery dataset. For Ethereum it is the \emph{extra\_data} field.} 

To get a more accurate picture of the distributions, the addresses were mapped to the entities that control them using three different methods: known identifiers, known addresses, and known clusters. Briefly, the first tries to match well-known pool tags with the block's identifier, the second cross-checks the block's reward addresses with known pool addresses, and the third uses known legal ties between pools and is applied after the first two have successfully attributed a block to a pool. The attribution data needed for this were obtained through public blockchain explorers\footnote{Such explorers include
\href{https://bitinfocharts.com}{BitInfoCharts}, \href{https://etherscan.io}{Etherscan}, and 
\href{https://www.blockchain.com}{blockchain.com}.} and community projects.\footnote{For example, the following 
project offers information about Bitcoin pools: 
\href{https://github.com/bitcoin-data/mining-pools}{github.com/bitcoin-data/mining-pools}.} If no match is found by any of these methods, then the block's address is considered a unique identity. Due to the pseudonymity of these systems and the fact that we cannot have attribution data for all addresses, it is of course possible that some of the ``unique identities'' are, in reality, controlled by the same entity. However, we do not expect this to be a  problem in our context, as the Nakamoto coefficient only looks at the ``biggest'' entities of the system, for which there typically exists plenty of information. 

Our data spanned three years, and the daily blocks mined range from 140 to 6,500 across the different ledgers. The number of unique entities mining these blocks ranged from 236 (Bitcoin) to 16021 (Ethereum). 
We count how many blocks were produced by each unique entity.

\begin{figure}
    \centering
    \begin{tikzpicture}
\draw[ultra thick, ->] (-1.5,0) -- (6,0);
\foreach \x in {-1.5, -0.5, 0.5, 1.5, 2.5, 3.5, 4.5}
\draw (\x cm,3pt) -- (\x cm,-3pt);

\draw[ultra thick] (-1.5,0) node[below=3pt,thick] {\footnotesize Day $1$} node[above=3pt] {};
\draw[ultra thick] (-0.5,0) node[below=3pt,thick] {\footnotesize Day $2$} node[above=3pt] {};
\draw[ultra thick] (0.5,0) node[below=3pt,thick] {\footnotesize Day $3$} node[above=3pt] {};
\draw[ultra thick] (1.5,0) node[below=3pt,thick] {\footnotesize Day $4$} node[above=3pt] {};
\draw[ultra thick] (2.5,0) node[below=3pt,thick] {\footnotesize Day $5$} node[above=3pt] {};
\draw[ultra thick] (3.5,0) node[below=3pt,thick] {$...$} node[above=3pt] {};

\draw[ultra thick] (3,0.15) node[above=3pt,thick] {$...$} node[above=3pt] {};

\draw [ black, ultra thick,decorate,decoration={brace,amplitude=5pt},
       ] (-1.5,0.5) -- (0.5,0.5)
       node [black,midway,above=4pt, align=center] {\footnotesize Window 1};
       
\draw [ black, ultra thick,decorate,decoration={brace,amplitude=5pt, mirror},
       ] (-0.5,-0.75) -- (1.5,-0.75)
       node [black,midway,below=4pt,align=center] {\footnotesize Window 2};

\draw [ black, ultra thick,decorate,decoration={brace,amplitude=5pt},
       ] (0.5,0.5) -- (2.5,0.5)
       node [black,midway,above=4pt, align=center] {\footnotesize Window 3};

    \end{tikzpicture}
        \caption{A sliding window with a 3-day granularity.}
    \label{fig:granularity figure}
\end{figure}
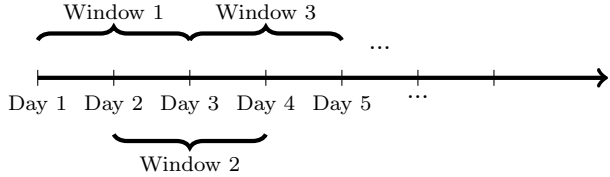

To translate this into time-series data, we use sliding window sampling where the data can be aggregated across $n$-days. Figure~\ref{fig:granularity figure} shows an example of this sliding window sampling with a 3-day sample window. Here, the first sample window consists of all blocks mined in days 1-3. The second sample window consists of all blocks mined in days 2-4, and so on. 
To calculate the Nakamoto coefficient for a given window, we count the number of blocks mined by each entity $i$ in the specific window. 
We then calculate $p_i$, the number of blocks mined by entity $i$ divided by the total number of blocks in that window, and use all the values $p_i$ to calculate the Nakamoto coefficient using the definition from Section~\ref{subsec: conceptual framework}.

We explore how changing the value of $n$, which we call the \emph{granularity} of the estimates, impacts statistical confidence.

\section{Results}
\label{sec:empirical results}
This section explores how statistical confidence in Nakamoto coefficient estimates is impacted by two core parameters, namely the significance level $\alpha$ and the granularity of the sample window.
It also compares uncertainty in estimates across different ledgers.
Throughout we chose sample windows that best illustrate the statistical methods that we developed.

Granularity impacts how sensitive the measurement is to fluctuations over time.
High granularity estimates aggregate over a shorter time period, which makes them more sensitive to changes in the distribution of mining power.
This helps participants to rapidly detect the potential for a 51\% attack.
However, high granularity also requires estimating with fewer blocks mined.
This lower $n$ increases the probability of observations deviating from the true distribution of mining power.
We explore the impact of aggregating daily, across 3-days, 7-days, 14-days, and 30-days. 
Ideally, researchers would use the lowest granularity that achieves a tolerable level of statistical confidence.

\begin{figure}
\begin{centering}
\includegraphics[width = \columnwidth]{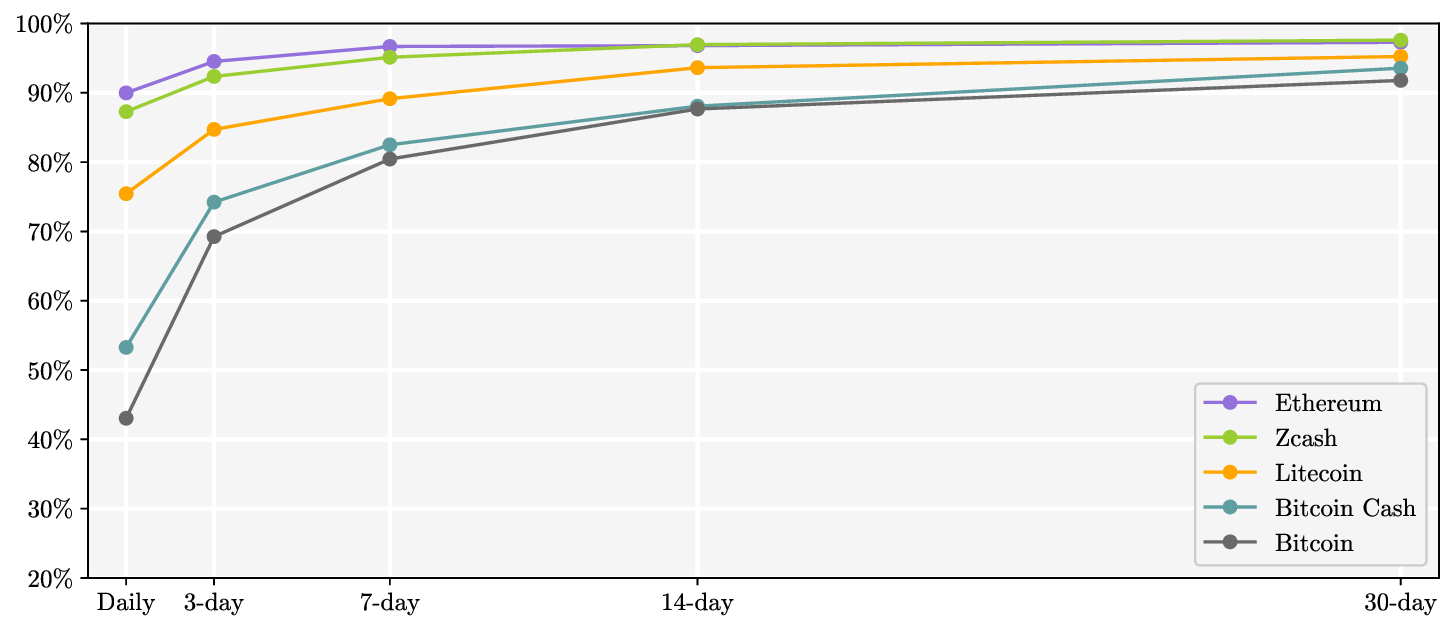}
\caption{How often a Nakamoto coefficient passes a hypothesis test at $\alpha=0.05$ with differing levels of granularity (based on our full dataset from 2018-2023).}
\label{fig: granularity alpha 0.05}
\end{centering}
\end{figure}

\begin{figure}

\begin{centering}
\includegraphics[width = \columnwidth]{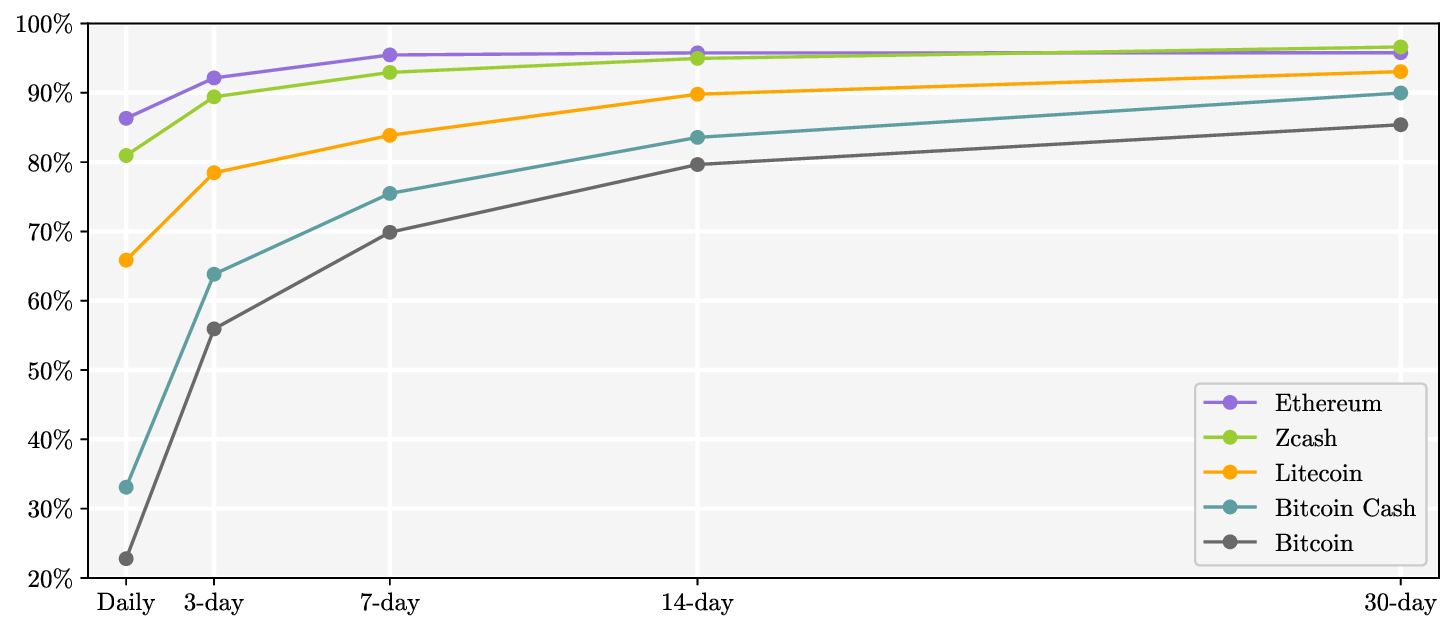}
\caption{With a smaller confidence level ($\alpha=0.01$), a smaller proportion of hypothesis tests are passed (based on our full dataset from 2018-2023).}
\label{fig: granularity alpha 0.01}
\end{centering}

\end{figure}

To evaluate granularity, we can test whether estimates with that granularity pass a hypothesis test at a given confidence level $\alpha$.
Figures \ref{fig: granularity alpha 0.05} and \ref{fig: granularity alpha 0.01} demonstrate the relationship between granularity and statistical confidence at significance levels $\alpha=0.05$ and $\alpha=0.01$ respectively. 
The most notable characteristic is that statistical uncertainty varies significantly across ledgers for the same granularity level.
This is driven by two factors: (i) blockchains with a higher throughput have more blocks in a given time period (more statistical power); and (ii) there is more statistical confidence in low values of Nakamoto coefficient because the distribution of blocks mined has less entropy.

Figure~\ref{fig: granularity alpha 0.05} demonstrates that when using the confidence level of $\alpha=0.05$ and a daily sample window, Nakamoto coefficient estimates for Bitcoin fail our hypothesis test more than 50\% of the time.
Bitcoin Cash shows similar results.
Ethereum and Zcash pass a statistical test 90\% of the time, even when using just a day's worth of block data. 
Aggregating blocks over 7 days creates a lot more statistical confidence, with all ledgers passing the hypothesis test over 80\% of the time (at a 7-day granularity, Bitcoin passes 80.45\% of hypothesis tests).
There are diminishing returns to increasing granularity beyond this point.
Even with monthly data, none of the ledgers pass out hypothesis test 100\% of the time.

We re-ran the same tests with a stricter significance level ($\alpha$).
Figure~\ref{fig: granularity alpha 0.01} shows how often each coin passes a hypothesis test with the significance level $\alpha=0.01$.
The same broad story holds---we see low statistical confidence in Bitcoin and Bitcoin Cash, whereas we are more confident in the estimates for Ethereum and Zcash.
Again, Bitcoin and Bitcoin Cash have lower throughput than the latter.
For this significance level, the diminishing returns set in for a 14-day sample window, after which the gains in statistical confidence are limited.

Thus far, we have explored statistical confidence in a single Nakamoto coefficient estimate calculated directly from the observed data.
However, it may make more sense to consider the \textit{potential} values of the Nakamoto coefficient.
The estimate calculated directly from the data is always the most plausible, however it is possible that higher and lower values may also be plausible if we cannot rule out these values at a given significance level.
In this way, our approach also provides us with insight into \textit{all} the potential values of the Nakamoto coefficient, as explained in Section \ref{subsec: conceptual framework}. For the sake of brevity, we focus on a 5\% significance level.\footnote{The Python code to perform this analysis with other values of $\alpha$ can be found in Appendix~\ref{appendix: code} and on this GitHub repository: \url{https://github.com/Blockchain-Technology-Lab/nc-statistical-confidence}.}

\begin{table*} 
\caption{Using the daily granularity and $\alpha=0.05$, we see a range of possible Nakamoto coefficients across ledgers. Red cells show us days where the possible Nakamoto coefficient was lower than the reported value, indicating a possible security risk.}
\label{table:nc ranges}
\centering
\adjustbox{max width=\textwidth}{
\begin{tabular}{ccc|cc|cc|cc|cc} 
\toprule

&\multicolumn{2}{c}{\thead{\textbf{Bitcoin}}}&\multicolumn{2}{c}{\thead{\textbf{Bitcoin Cash}}} & \multicolumn{2}{c}{\thead{\textbf{Ethereum}}} & \multicolumn{2}{c}{\thead{\textbf{Litecoin}}} & \multicolumn{2}{c}{\thead{\textbf{Zcash}}} \\ 

& Direct & \multicolumn{1}{c}{Possible} & Direct & \multicolumn{1}{c}{Possible} & Direct & \multicolumn{1}{c}{Possible} & Direct & \multicolumn{1}{c}{Possible} & Direct & \multicolumn{1}{c}{Possible} \\
\midrule
Jan 1, 2019 & \cellcolor{marycolor}4 &\cellcolor{marycolor} 3-4 & \cellcolor{marycolor}2 & \cellcolor{marycolor}1-2 &\cellcolor{marycolor} 3 & \cellcolor{marycolor}2-3 & 3 & 3 & 2 & 2\\

Jan 2, 2019 & \cellcolor{marycolor}4 & \cellcolor{marycolor}3-4 & \cellcolor{marycolor}2 & \cellcolor{marycolor}1-2 & 2 & 2 & 3 & 3 & 2 & 2\\

Jan 3, 2019 & \cellcolor{marycolor}4 & \cellcolor{marycolor}3-4 & 2 & 2 & 2 & 2 &\cellcolor{marycolor} 4 &\cellcolor{marycolor} 3-4 &\cellcolor{marycolor} 3 & \cellcolor{marycolor}2-3\\

Jan 4, 2019 &\cellcolor{marycolor} 4 & \cellcolor{marycolor}3-4 & 2 & 2 & 2 & 2 &\cellcolor{marycolor} 4 &\cellcolor{marycolor} 3-4 & 3 & 3\\
Jan 5, 2019 & \cellcolor{marycolor}3 & \cellcolor{marycolor}2-3 & \cellcolor{marycolor}2 & \cellcolor{marycolor}1-2 & 2 & 2 & 3 & 3 &\cellcolor{marycolor} 3 &\cellcolor{marycolor} 2-3 \\
Jan 6, 2019 & \cellcolor{marycolor}4 & \cellcolor{marycolor}3-4 & 2 & 2 & 2 & 2 & 3 & 3 & \cellcolor{marycolor}3 & \cellcolor{marycolor}2-3\\

Jan 7, 2019 & \cellcolor{marycolor}4 &\cellcolor{marycolor} 3-4 & 2 & 2 & 2 & 2 &\cellcolor{marycolor} 4 & \cellcolor{marycolor}3-4 &\cellcolor{marycolor} 3 &\cellcolor{marycolor} 2-3\\

Jan 8, 2019 & 4 & 4 & 2 & 2 &\cellcolor{marycolor} 3 &\cellcolor{marycolor} 2-3 &\cellcolor{marycolor} 4 &\cellcolor{marycolor} 3-4 & 3 & 3\\

Jan 9, 2019 & 3 & 3 &\cellcolor{marycolor} 3 &\cellcolor{marycolor} 2-3 & 2 & 2 &\cellcolor{marycolor} 4 &\cellcolor{marycolor} 3-4 &\cellcolor{marycolor} 3 &\cellcolor{marycolor} 2-3\\
Jan 10, 2019 & 3 & 3 & 2 & 2 & 2 & 2 &\cellcolor{marycolor} 4 & \cellcolor{marycolor}3-4 & 2 & 2\\
Jan 11, 2019 &\cellcolor{marycolor} 4 & \cellcolor{marycolor}3-4 &\cellcolor{marycolor} 3 & \cellcolor{marycolor}2-3 & 2 & 2 & 4 & 4 & \cellcolor{marycolor}3 &\cellcolor{marycolor} 2-3\\

Jan 12, 2019 & \cellcolor{marycolor}4 &\cellcolor{marycolor} 3-4 & 2 & 2 & 2 & 2 &\cellcolor{marycolor} 4 &\cellcolor{marycolor} 3-4 &\cellcolor{marycolor} 3 &\cellcolor{marycolor} 2-3\\
Jan 13, 2019 & 3 & 3 & 2 & 2 & 2 & 2 &\cellcolor{marycolor} 4 &\cellcolor{marycolor} 3-4 & 3 & 3\\
Jan 14, 2019 & 3 & 3 & 2 & 2 & 2 & 2 & 3 & 3 & 3 & 3\\

\bottomrule
\end{tabular}}
\end{table*}

Table~\ref{table:nc ranges} compares the direct value of the Nakamoto coefficient to the range of possible values using our method based on a daily granularity.
Across the 2-week period in 2019, 44\% of the direct values could plausibly be lower.
At a 5\% significance level, we could not reject the possibility that a single entity held the majority of the Bitcoin Cash mining power on the \nth{1}, \nth{2} and \nth{5} of January 2019.
For the other ledgers, there are no days where a single entity might plausibly control a majority of the mining power. 
However, for both Bitcoin and Zcash, there were days where 2 entities could plausibly have held the majority of the mining power.

More generally, we find that the direct values consistently under-estimate how centralized these blockchain systems are.
In all cases in Table~\ref{table:nc ranges}, the possible values are lower than the direct value.
This result is not because we do not test whether higher values of the Nakamoto coefficient are possible.
Indeed, when we use stricter significance levels, we find possible values that are higher than the direct value, which can be seen in Table~\ref{table: nc and alpha}.
However, there is a structural bias in that direct estimates of the Nakamoto coefficient \emph{under-estimate} how centralized blockchain systems are.

\begin{figure}
\centering
\includegraphics[width = \columnwidth]{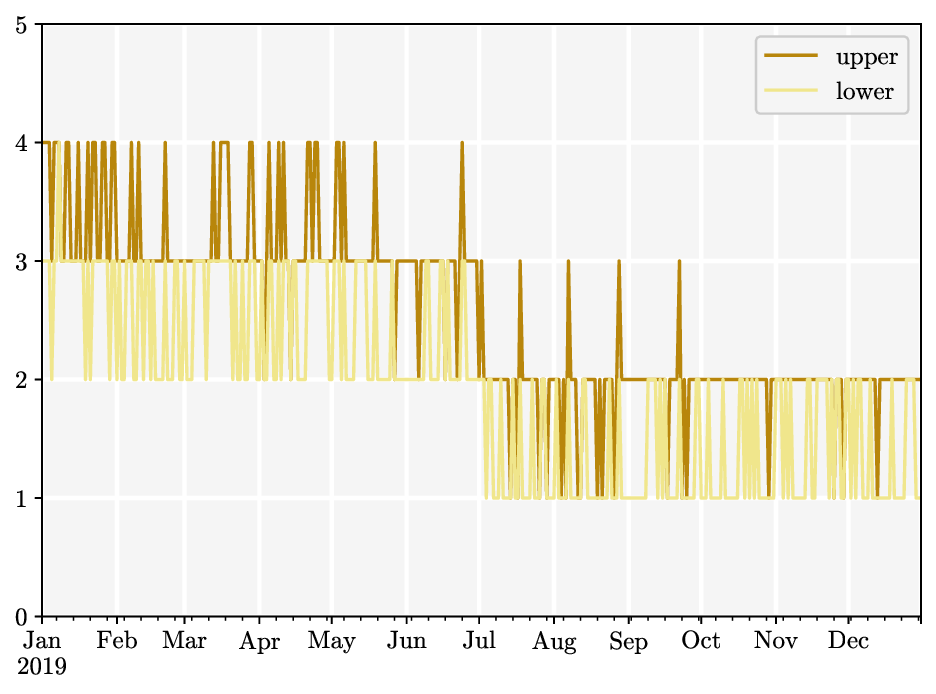}
\caption{The range of Nakamoto coefficient estimates (upper and lower) for Bitcoin with $\alpha=0.05$.}
\label{fig: bitcoin range 1 year}
\end{figure}

Zooming out to a longer time frame, Figure~\ref{fig: bitcoin range 1 year} shows how the range of possible Nakamoto coefficients unfolded throughout 2019 for Bitcoin. 
There is rarely statistical support to say there is exactly one true estimate, and instead the Nakamoto coefficient should be reported as a range.
The summer of 2019 appears to be a transitional period in which the Bitcoin ecosystem shifted towards more centralization in the distribution of mining power. In particular, there were days in which a Nakamoto coefficient estimate of $1$ was plausible.
Notably, a low value for the lower estimate should be seen as a potential security risk.

For Table~\ref{table:nc ranges} and Figure~\ref{fig: bitcoin range 1 year}, we fixed $\alpha=0.05$. 
Table~\ref{table: nc and alpha} explores how the Nakamoto coefficient estimates for Bitcoin change when we vary the confidence level, again for the 2-week period (Jan 1 2019 - Jan 14, 2019).
Applying the strict confidence level of 0.1\%, plausible values range from 2 to 5 on the \nth{6} of January.
We cannot say what level of statistical confidence is appropriate, but it clearly makes a difference when estimating the Nakamoto coefficient.

\begin{table} 
\caption{With varying levels of $\alpha$, the range of
the true Nakamoto coefficient for Bitcoin once again varies.}
\label{table: nc and alpha}
\centering
\adjustbox{max width=\textwidth}{
\begin{tabular}{ccccccc} 
\toprule

$\alpha = $& &0.001 & 0.01 & 0.025 & 0.05 & 0.1\\
\midrule
Jan 1, 2019& \textbf{4}  & 3-4 & 3-4 & 3-4 & 3-4 & 3-4\\
Jan 2, 2019 & \textbf{4}& 3-4 & 3-4 & 3-4 & 3-4 & 3-4 \\
Jan 3, 2019& \textbf{4} & 2 -4 & 3-4 & 3-4 & 3-4 &3,4\\
Jan 4, 2019 &  \textbf{4} & 3-4 & 3-4 & 3-4 &3-4 &3-4\\
Jan 5, 2019& \textbf{3} & 2-3 & 2-3 & 2-3 &2-3 &3 \\
Jan 6, 2019 & \textbf{4} &  2 - 5 & 3-4 &3-4 &3-4 &3-4\\
Jan 7, 2019& \textbf{4} & 3-4 &3-4 & 3-4& 3-4 &3-4 \\
Jan 8, 2019& \textbf{4} &  3-5 & 3-4 &4 &4 &4 \\
Jan 9, 2019& \textbf{3} &  2- 4 & 3-4 &3-4 &3 &3 \\
Jan 10, 2019 & \textbf{3} & 2- 4 & 2-3 &3 &3 &3\\
Jan 11, 2019 & \textbf{4}&  2- 4 & 3-4 &3-4 &3-4 &3-4 \\
Jan 12, 2019 & \textbf{4} &  2- 4 & 3-4 & 3-4 &3-4 &3-4\\
Jan 13, 2019 & \textbf{3} & 2- 4 & 2-3 & 3 &3 &3\\
Jan 14, 2019 & \textbf{3} & 2- 4 & 2-3 & 2-3 &3 &3\\

\bottomrule
\end{tabular}}
\end{table}

\section{Related Work}
\label{sec:litreview}
The Nakamoto coefficient is regularly used to summarize mining power distribution~\cite{quantifying, demonitor, measuringdec, evolution, individual, 9488812, romiti2019deep, LIN2022112620, grandjean2023ethereum}. 
Other works use the Nakamoto coefficient to capture other layers of the blockchain, namely governance token and/or token distribution~\cite{jensen2021decentralized, 9854972, fritsch2022analyzing}.
We focus on a handful of articles that illustrate why statistical confidence matters and how to use our framework.

In 2017, Srinivasan and Lee~\cite{quantifying} introduced the Nakamoto coefficient. In doing so, they calculated it for Bitcoin and Ethereum on a single day (\ie daily granularity). Although the specific date is not mentioned, we are led to believe it is the 24-hour period preceding publication on 28 July 2017.
In this inaugural paper, Srinivasan and Lee reported that Bitcoin had a Nakamoto coefficient of 5. However, even with the standard significance level ($\alpha = 0.05$), we could not rule out the possibility that the Nakamoto coefficient could have been 4. When investigating the source of Srinivasan and Lee's data~\cite{quantifying}, the dashboard warns that ``our analysts have found that weekly numbers are a better representation of the underlying power, because they are less sensitive to mining randomness''~\cite{blockchain}, but they do not support this statement with statistical analysis.
For their sample window, Ethereum likely had sufficient statistical confidence (see Figure~\ref{fig: granularity alpha 0.05}).\footnote{It was a PoW blockchain in 2017, and so they faced the problem of statistical uncertainty.}

Lin \textit{et al.} \cite{measuringdec} measure the same two ledgers, but use three different levels of granularity.
Two of these (weekly and monthly) are likely appropriate, although reporting this via statistical tests would increase our confidence. 
However, they also report on estimates using a daily granularity, which is not appropriate for Bitcoin. 

Liu \textit{et al.} \cite{demonitor} also use three levels of granularity (daily, weekly, and monthly) to summarize data for Dogecoin and Ethereum Classic.
Both blockchains have high throughput and, thus, likely have sufficient statistical confidence for all granularities.
Nevertheless, reporting on statistical tests using our framework would support this choice.

Compajola \textit{et al.} \cite{evolution} use a weekly granularity for a range of ledgers.
This weekly granularity conforms with our recommendations, although we did not check whether it is appropriate for Monacoin or Feathercoin.
Finally, Li \textit{et al.}~\cite{individual} used a daily granularity for Steem and Ethereum.
It is unclear whether this granularity was appropriate for Steem, which we did not investigate.
For new ledgers, researchers need to run their own statistical analysis using the framework introduced in Section~\ref{sec:approach} and the code published in the Appendix.

\section{Discussion} \label{sec:discussion}
This section discusses implications and future work.\\

\textbf{Implications}
It is natural to ask whether reporting statistical confidence matters. 
As discussed, most researchers calculate the Nakamoto coefficient by assuming the observed distribution of mined blocks is equal to the true distribution of mining power.
In their defense, this estimate is the most plausible value, even though there may be other plausible values.
All of the possible values in Table~\ref{table:nc ranges} include the value that most researchers calculate by default.

Statistical uncertainty matters when it has security implications.
For example, the default approach might lead researchers to estimate a Nakamoto coefficient of 2, even though 1 is a plausible value.
This would mean there is a non-negligible probability that one entity can launch a safety attack.
In this way, the direct approach---universally used in industry and academia---can create a false sense of security.
In particular, we have shown that the direct approach appears to have a bias towards under-estimating centralization, at least in the blockchains and time periods that we studied.

One potential criticism of our work is that we run too many statistical tests.
Ioannidis~\cite{ioannidis2005most} made the influential argument that most scientific findings are false because the significance level $\alpha=0.05$ is vulnerable to various publication biases when researchers run many statistical tests and selectively report on the significant ones.
In the same way, one might argue that our study ran thousands of statistical tests, and so it is unsurprising that the direct values lacked statistical support some of the time.
We accept that some estimates lacking support is inevitable. However, it happens too often for some levels of granularity.
For example, Figure~\ref{fig: granularity alpha 0.05} shows that most Nakamoto coefficient estimates for Bitcoin do not have statistical support based on a daily granularity.

Going forward, one solution is for researchers to report statistical uncertainty using the framework introduced in this paper.
The Nakamoto coefficient can be reported as a range of plausible values (see Table~\ref{table: nc and alpha}).
It is much harder to do the equivalent for other metrics (see Appendix~\ref{sec:othermetrics}) because of their complexity.
However, we acknowledge that not all authors will report a range of values.
For this reason, we reluctantly make the recommendation that a granularity of at least a week is used when estimating the Nakamoto coefficient for PoW blockchains.
Such values achieve most of the reduction in statistical uncertainty.
However, it is possible that systems exist that require a a longer sample window, such as those with particularly low throughput.
\vspace{0.1cm}

\noindent\fbox{%
    \parbox{\columnwidth}{%
        \textbf{(Reluctant) Recommendation}\\
        If researchers do not report a range of values, they should calculate Nakamoto coefficient values using a granularity of at least 7 days.
    }%
} \vspace{0.1cm}

\textbf{Future Work}
There are additional directions for methodological contributions. 
One direction is to identify statistical tests for other metrics that are used to quantify blockchain decentralization, such as the Shannon entropy, Gini coefficient, and/or 
Herfindahl–Hirschman index (HHI).
Appendix~\ref{sec:othermetrics} begins to explore the mathematical relationship between these metrics and the Nakamoto coefficient, specifically for the case where the coefficient is 1.
Future work should generalize this formal analysis for arbitrary values and also the relationship between other metrics, for example Gini and HHI.

For the Nakamoto coefficient, we defined granularity as a fixed time window (\eg 1-day or 7-day). It is possible to instead base each calculation on a fixed number of blocks mined.
This would ensure every estimate has sufficient statistical power.
However, it would also mean that blockchains with high throughput, say Ethereum, would be more sensitive to temporal variations than blockchains with lower throughput, say Bitcoin.
For example, Bitcoin would contribute 1000 new blocks approximately once a week, whereas Ethereum does this every 4 hours.

More fundamentally, statistical uncertainty is just one potential bias or inaccuracy related to decentralization estimates.
Attribution uncertainty is another source of error.
For example, a system may become centralized if two entities secretly coordinate actions (or are in fact the same entity).
Similar hazards may occur when individual miners offer significant power in parallel to
multiple, seemingly independent pools~\cite{romiti2019deep}.
Additionally, block attribution tags, such as Bitcoin's \textit{coinbase} parameter, are filled by pools on a voluntary basis. In particular, although pools tend to claim blocks as their own by adding their attribution information in them, this information cannot be independently verified, so some level of trust is necessary, i.e., assuming that the information is correct.
In summary, block attribution is typically done in an ad-hoc manner, so the research community should build robust techniques and datasets to address the problem of mapping out the ecosystem of entities and their relationships.

\section{Conclusion} \label{sec:conclusion}
Block mining is a stochastic process, with success determined by each entity's mining power.
For PoW blockchains, the distribution of mining power cannot be read directly from the blockchain, and must instead be inferred from the number of blocks mined.
We introduced a framework to quantify this statistical uncertainty.

We showed that aggregating blocks daily can lead to considerable uncertainty when estimating the Nakamoto coefficient.
For example, estimates for Bitcoin from 2018--2023 fail a hypothesis test more than half the time when using a day's worth of block data.
Blockchains with higher throughput had less statistical uncertainty.
We also showed that the existing approach to calculating the Nakamoto coefficient consistently underestimates the centralization of mining power, which provides a false sense of security regarding the risk of a 51\% attack.

Looking forward, a granularity of 7-14 days appears to achieve the best balance between statistical power and sensitivity to temporal fluctuations in the distribution of mining power, at least among the blockchains that we studied.
We also recommend that authors report a range of possible Nakamoto coefficient values, which have statistical support at a particular significance level $\alpha$.
To allow researchers to run hypothesis tests and calculate this range, we produced the relevant code in Python, which is publicly available on GitHub\footnote{\url{https://github.com/Blockchain-Technology-Lab/nc-statistical-confidence}} and can be used by anyone.

\section*{Acknowledgements}
This work was supported by Input Output (iohk.io) through their funding of the Edinburgh Blockchain Technology Lab.

\ifccs
\else

\bibliographystyle{alpha}

\fi

\bibliography{references}

\appendix

\section{Other Metrics} \label{sec:othermetrics}
Given our framework is tailored to the Nakamoto coefficient, we explored the mathematical relationship between the Nakamoto coefficient and other popular decentralization metrics. In particular, we focused on whether other metrics can identify distributions of mining power with the highest security risk. This occurs when the Nakamoto coefficient is 1, which means a single entity holds more than 50\% of the mining power. We explore possible values of the other decentralization metrics when the Nakamoto coefficient is 1.
 
Our analysis reveals that, in this setting, the HHI must be greater than 2500, Shannon entropy can be any value, and the Gini coefficient is 0.5 or more if n is sufficiently high.

\subsection{Herfindahl–Hirschman Index (HHI)}
The HHI is  the sum of the squares of each entity's percentage of market control: 
\[{\displaystyle HHI=\sum _{i=1}^{N}(s_{i})^{2}}\] It ranges from 0 to 10,000, where higher values indicate a more concentrated market, and therefore a system which is more vulnerable to a safety attack. If we assume that 1 person holds the majority of the resources, then the HHI will be minimized when the remaining resources are held in equal share ($s_i = \frac{1-s_1}{n-1}$, where $s_i$ is the share of resources held by the $i$-th entity). Let $s_1$ be the share of the market controlled by the most powerful entity. Then,
\[\text{HHI} = s_1^2 + s_2^2 + s_3^2 + \ldots\]
\[\text{HHI} = s_1^2 + \left( \frac{1-s_1}{n-1} \right)^2 + \left( \frac{1-s_1}{n-1} \right)^2 + \ldots\]
\[\text{HHI} = s_1^2 + (n-1)\left( \frac{1-s_1}{n-1} \right)^2\]
\[\text{HHI} = s_1^2 + \frac{(1-s_1)^2}{n-1}\]
This value is lowest for very large $n$.
Therefore,
\[\lim_{n\to\infty} \text{HHI} = s_1^2.\]
So, if the Nakamoto coefficient is 1 ($s_1 > 0.50$),
\[ \text{HHI} > 2500.\]
This means that given an HHI of less than or equal to 2500, we know that the Nakamoto coefficient must be greater than 1. 

\subsection{Shannon entropy}
Formally, Shannon entropy is defined as: 
\[\mathrm{H} (X):=-\sum_{x\in 
{\mathcal{X}}}p(x)\log p(x)\] where $X$ is a discrete random variable that 
takes values in $\mathcal{X}$ and $p(x)$ is the probability of an outcome $x 
\in \mathcal{X}$. 
A system with high entropy is a decentralized system. Once again we consider a setting with a Nakamoto coefficient of 1 where $p_1$ is the proportion of resource ownership by the majority party. Once again, we know that entropy will be maximized if the remaining $n-1$ parties hold the remaining resources in equal share, $p_i = \frac{1-p_1}{n-1}$. 
\\
\resizebox{.9\hsize}{!}{$\text{E} = -(p_1\log_2p_1 + \frac{1-p_1}{n-1}\log_2\left(\frac{1-p_1}{n-1}\right) + \frac{1-p_1}{n-1}\log_2\left(\frac{1-p_1}{n-1}\right) + \ldots)$}
\[\text{E} = -p_1log_2p_1 - (n-1)\frac{1-p_1}{n-1}\log_2\left(\frac{1-p_1}{n-1}\right)\]
\[\text{E} = -p_1log_2p_1 - (1-p_1)\log_2\left(\frac{1-p_1}{n-1}\right)\]
\[\text{E} = -p_1log_2p_1 - (1-p_1)\log_2(1-p_1)+ (1-p_1)\log_2(n-1)\]
Then,
\[\lim_{n\to\infty} \text{E} = (-1+p_1)(-\infty).\]
Therefore, if the Nakamoto coefficient is 1 ($p_1 > 0.5$)\footnote{We also assume here that majority party does not own 100\% of the resources.}, 
\[\lim_{n\to\infty} \text{E} > \infty.\]

So, with a high enough number of participants, we could observe a high value of entropy, but was due to having a large population rather than actual security.

This means regardless of the observed value for Shannon entropy, it is possible that the Nakamoto coefficient is 1.

\subsection{Gini Coefficient}
The Gini coefficient is defined as
\[\text{G} = \frac{\text{A}}{\text{A}+\text{B}},\]
where B is the area under the Lorenz curve and A is the area between the line of equality and the Lorenz curve. A Gini coefficient of 0 indicates perfect equality, and a coefficient of 1 indicates perfect inequality. Since there are no negative incomes, then $A+B = 0.5$ \cite{EXCEL}, and
\[\text{G} = 2\text{A} = 1 - 2\text{B} \]
. Once again, we will assume that one party holds a majority proportion of $x_1$ of the resource and the other $n-1$ hold $\frac{1-x_1}{(n-1)}$ in equal share, then
\[\text{B} = \frac{x_1}{n} + \left(\sum_{i=1}^{n-1}\frac{1-x_1}{n-1}\times i \times \left(\frac{1}{n}\right)\right)\]
\[\text{B} = \frac{x_1}{n} + \frac{1- x_1}{2} \]
If $x_1 \geq 0.5$, then
\[\text{B} \geq \frac{1}{2n} + \frac{1}{4} \]
So,
\[\text{G} \geq 1-2\left(\frac{1}{2n}+\frac{1}{4}\right)\]
and, 
\[\lim_{n\to\infty} \text{G} \geq \frac{1}{2}.\]

Most fundamentally, a system in which two parties held 50\% of the mining power would have a Gini coefficient of 0, indicating decentralization, but would be, in reality, a highly centralized system. 
However, we showed that for a sufficiently large number of participants $n$, a Gini coefficient less than $\frac{1}{2}$ cannot hide a safety attack (unlike with entropy).

\section{Code}
\label{appendix: code}
We include code snippets here that can be used to calculate the Nakamoto coefficient and also the range of possible values at a given significance level $\alpha$. The full code that was used to produce our results can be found on this public GitHub repository: \url{https://github.com/Blockchain-Technology-Lab/nc-statistical-confidence}. We have also included some sample data there that can be used as input.

\lstset{ 
language=Python, 
tabsize=2, 
showspaces=false, 
showstringspaces=false, 
backgroundcolor=\color{listgray}, 
float=[htb], 
captionpos=b, 
basicstyle=\footnotesize, 
frame=tbrl, 
frameround=tttt, 
numbers=left, 
numberstyle=\tiny, 
numberblanklines=false, 
linewidth=\textwidth}

\begin{figure*}
\begin{lstlisting}[breaklines, caption = Python code to compute the Nakamoto coefficient.]
import pandas as pd

def compute_nakamoto_coefficient(row):
    """
    :param row: series of blocks mined by each distinct entity
    :returns: nakamoto coefficient for the given row
    """
    total_blocks = sum(row)
    nc, power_ratio = 0, 0
    if total_blocks > 0:
        for blocks in row.sort_values(ascending=False):
            nc += 1
            power_ratio += blocks / total_blocks
            if power_ratio > 0.5:
                break
    return nc


def compute_nakamoto_coefficients(df):
    """
    :param df: dataframe of blocks mined, with
        dates in rows and distinct entities in columns
    :returns: dataframe of nakamoto coefficient, with
        dates in rows and corresponding values of nc in columns
    """
    nc_series = df.apply(lambda row: compute_nakamoto_coefficient(row), axis=1)
    nc_df = pd.DataFrame( {'nc': nc_series}, index=df.index)
    return nc_df
\end{lstlisting}
\end{figure*}

\begin{figure*}
\begin{lstlisting}[breaklines, caption = Python code to find the range of possible Nakamoto coefficients.]
from scipy.stats import binomtest
import pandas as pd

def find_nc_range(df, nc_df, alpha=0.05):
    """
    :param df: dataframe of blocks mined, with
        dates in rows and distinct entities in columns
    :param nc_df: dataframe of nakamoto coefficient, with
        dates in rows and corresponding values of nc in columns
    :returns: dataframe of range of nakamoto coefficient values, with
        dates in rows and lower, upper nakamoto coefficient in columns
    """
    lower, upper = [], []
    for date in df.index:
        total_blocks = df.loc[date].sum(axis=0) 
        coeff = nc_df['nc'].loc[date]
        coeffp, coeffq = coeff, coeff
        if total_blocks > 0:
            sorted_df = df.loc[date].sort_values(axis=0, ascending=False)
            successes = sorted_df.nlargest(coeff).sum()
            p = binomtest(k=successes, n=total_blocks, p=0.5, alternative='greater').pvalue
            if p > alpha:
                while p > alpha:  # upper
                    coeffp += 1
                    upper_sorted = df.loc[date].sort_values(axis=0, ascending=False)
                    successes = int(upper_sorted.nlargest(coeffp).sum())
                    p = binomtest(k=successes, n=total_blocks, p=0.5, alternative='greater').pvalue
                coeffp -= 1
            q = binomtest(k=successes, n=total_blocks, p=0.5, alternative='less').pvalue
            if q > alpha:
                while q > alpha:  # lower
                    coeffq -= 1
                    lower_sorted = df.loc[date].sort_values(axis=0, ascending=False)
                    successes = int(lower_sorted.nlargest(coeffq).sum())
                    q = binomtest(k=successes, n=total_blocks, p=0.5, alternative='less').pvalue
                coeffq += 1
        lower.append(coeffq)
        upper.append(coeffp)
    result = pd.DataFrame({'lower': lower, 'upper': upper}, index=df.index)
    return result
\end{lstlisting}
\end{figure*}

\end{document}